# Lawmaps: Enabling Legal AI development through Visualisation of the Implicit Structure of Legislation and Lawyerly Process

Scott McLachlan[a,b,c,1], Evangelia Kyrimi[b], Kudakwashe Dube[c,d], Norman Fenton[b] and Lisa C Webley [a]

[a] *Birmingham Law School, Birmingham University, Birmingham (UK)*
[b] *Risk and Information Management Group, Queen Mary University of London (UK)*
[c] *Health informatics and Knowledge Engineering Research Group (HiKER)*
[d] *School of Fundamental Sciences, Massey University, Palmerston North (NZ)*

**Abstract.** Modelling that exploits visual elements and information visualisation are important areas that have contributed immensely to understanding and the computerisation advancements in many domains and yet remain unexplored for the benefit of the law and legal practice. This paper investigates the challenge of modelling and expressing structures and processes in legislation and the law by using visual modelling and information visualisation (InfoVis) to assist accessibility of legal knowledge, practice and knowledge formalisation as a basis for legal AI. The paper uses a subset of the well-defined Unified Modelling Language (UML) to visually express the structure and process of the legislation and the law to create visual flow diagrams called lawmaps, which form the basis of further formalisation. A lawmap development methodology is presented and evaluated by creating a set of lawmaps for the practice of conveyancing and the Landlords and Tenants Act 1954 of the United Kingdom. This paper is the first of a new breed of preliminary solutions capable of application across all aspects, from legislation to practice; and capable of accelerating development of legal AI.

## 1. Introduction

Visual representation of law and legal processes has potential to systematise decision-making in a way that mitigates risk and improves efficiency for lawyers, and lessens confusion and misunderstanding for their clients (Leiman, 2016); both of which could decrease costs, especially those incurred relitigating matters (Fang, 2014). In a system which is largely client funded this means there is potential for greater access to good quality legal assistance that a larger portion of the community can afford. While historic visual approaches like Wigmore Charts (Fig. 1) and Beardsley Diagrams were proposed for analysis of trial evidence (Reed, Walton & Macagno, 2007)[2], and newer methods like Araucaria Diagrams and Bayesian Networks (Fig. 2) have been used as visual steps to computational representation of legal argument in construction of AI (Reed et al, 2007; Fenton, Neil & Lagnado, 2013), the use of process or flow diagrams to describe the inner workings of legislation or the lawyerly process remains rare (McLachlan & Webley, 2020). This means that while diagrams and AI exist that exemplify a particular argument type, individual case or court judgment, there are very few exposing broader legal issues or categories of cases.

[1] Corresponding Author: Dr Scott McLachlan, Birmingham Law School, University of Birmingham, Edgbaston, Birmingham, B15 2TT, United Kingdom; E-mail: s.mclachlan@bham.ac.uk
[2] The Wigmore Chart is a graphical analysis method for trial evidence developed by John Wigmore in the early 1900s, while Beardsley proposed a number of visualisation approaches for argument mapping during the 1950's.

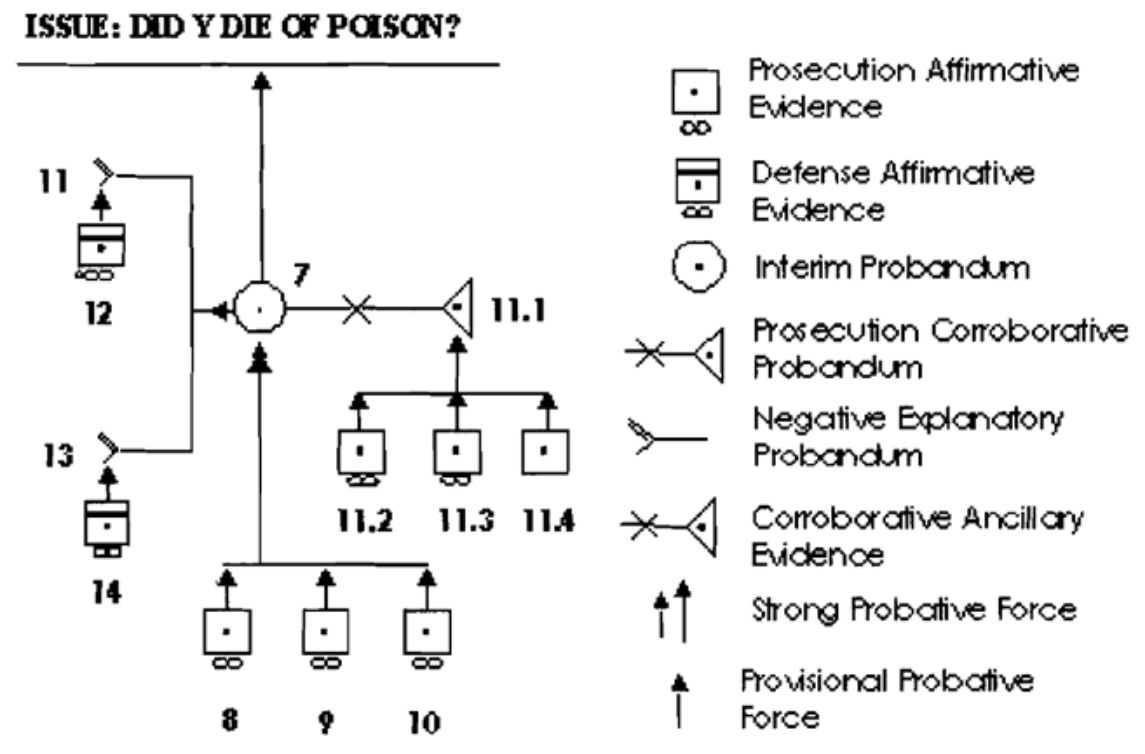

**Figure 1.** Wigmore diagram reproduced from Reed et al (2007).

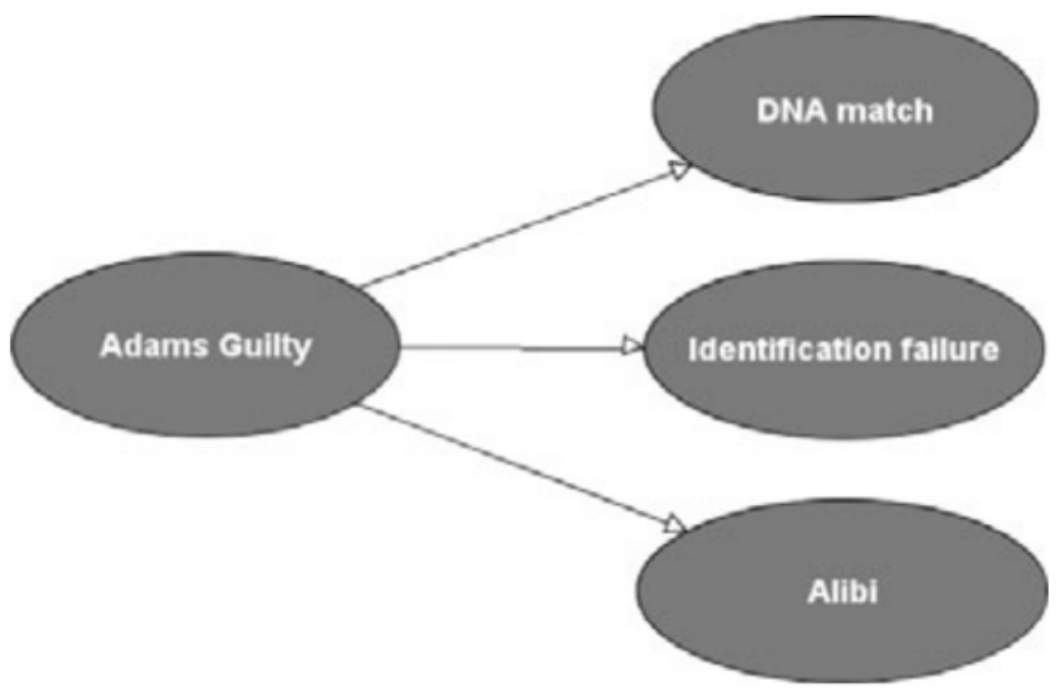

**Figure 2.** Bayesian Network for the case of *R v Adams* reproduced from Fenton et al (2013).

Visualisations of law and justice can be an important component for how lawyers and society more widely engage with the legal system. Legislation is infused with structure, and not just the *visible structure* of chapters, headings and statutory sections, but also an *implicit structure* that only becomes apparent when different sections and subsections interact as that legislation is analysed or when they are applied to a particular set of circumstances. The complex, verbose and abstruse nature of legislation means that it is difficult for those not trained in the law to comprehend the structure and meaning of legislation.

The systematic collation of interactions between legal provisions, processes and decision points could provide a formidable tool to support lawyer's decision-making and a powerful representation of the law and legal system. Information visualisation (infovis) provides a mapping between discrete data and visual representations in order to improve memory, cognition and comprehension of otherwise abstract information (Huron, 2014; Manovich, 2011). Infovis reduces the variability of what is specific to individual instances of data or an activity to visually represent only the core characteristics in order to reveal the patterns and structures inherent to that data or activity. The process of developing infovis draws on two core principles (Manovich, 2011):

- The *reduction* of complex data and relationships to graphical primitives such as points, lines, curves and simple geometric shapes, and;
- Use of *spatial variables* such as position, size, shape, colour and movement to represent key differences in data and reveal patterns.

Lawyers make countless decisions in the course of daily practice. Decisions about the relevant legal provisions that apply to a given matter, the scope of advice they give, the investigations they undertake on behalf of their client, and the context of submissions they draft. The vast array of complex legislation, practice rules and precedent all serve to make legal decision-making a challenging task (James, 2007; Johnson, 1999; Mulvenna & Hughes, 1995). Lawyers adhere to relatively traditional ways of doing things, even as professionals in other domains adopt new technologies (Becerra, 2018). While there have been numerous calls to develop approaches and train lawyers using infovis (Koch, 2010; Leiman, 2016; McCloskey, 1998), these calls presently remain unanswered (Leiman, 2016; McLachlan et al, 2020).

Two problems are to be addressed in this work. The first is development of an infovis approach for the legal domain that can equally represent both the implicit structure of legislation and the stepwise nature of lawyerly processes. These visualisations are referred to as *lawmaps*. The second is to ensure that lawmaps are sufficiently democratised for adoption not just by experienced professionals, but also by law students and the general public. The goal is to reduce the apparent complexity of law and make lawmaps, and lawvis, ubiquitous tools that can enable law professionals, students and lay-people to visualise everyday legal activities.

## 2. The role of lawmaps in the pathway to legal AI

Early approaches to developing legal AI focused solely on understanding and modelling legal argument (Bench-Capon, 1997; Rissland, 1990). More recent projects have developed formulaic or rule-based approaches for performing routine tasks including predictive coding processes for e-Discovery, ranking of precedents during legal research, and predetermined routine document generation (Becerra, 2018). While there have been a number of forays into predictive analytics these have usually sought to provide a limited prediction as to the potential outcome if a matter proceeds to trial, or the decision a particular judge might be expected to make based on past rulings (Becerra, 2018). Issues have been raised with other predictive approaches that have drawn critical attention in the media, with claims that systems such as COMPAS[3] and Predpol[4] are racist and prejudicially biased. However, these applications of AI more correctly exist within the policing and justice domains and are not routinely used by lawyers. For this reason, while their striking ethical issues should be foremost in the minds of all who are developing intelligent solutions, we consider them to exist outside the scope of AI for legal practice.

Real-world problems generally comprise multiple related but uncertain variables and data (Constantinou et al, 2016). Developing effective AI involves two major tasks: (1) determining the structure; and (2) specifying the computational parameters and values (Kyrimi et al, 2020). It can be difficult to determine the most appropriate inputs when developing a new AI. However, one approach that has shown success is when data is combined with expert knowledge (Constantinou et al, 2016; Kyrimi et al, 2020). The process of deriving knowledge from experts, or *expert elicitation* (Constantinou et al, 2016; Kyrimi et al, 2020), allows for interdisciplinary collaboration between those developing the AI and experienced practitioners from the domain under investigation. The goal should not be wholesale replacement of the expert practitioner. Rather, the goal should be to develop tools that can aid or enhance their practice, and reproduce elements of their expertise that may assist the untrained or inexperienced to identify when and why they have an issue that requires expert input.

[3] The Correctional Offender Management Profiling for Alternative Sanctions (COMPAS) tool provides a recidivism risk scoring for each offender based on a range of factors that includes the offenders history, violent tendencies, associations, financial status, family history, attitudes and personality. A number of recent research publications have received high-profile media attention for demonstrating that while COMPAS is not explicitly told the ethnicity of the offender, persons of colour are repeatedly shown to receive scores that suggest higher rates of potential for recidivism than white offenders who committed equivalent crimes - even a white offender with a history of significant offending and prior incarceration.

[4] Predpol is a tool that analyses crime data to identify spot patterns of criminal behaviour. Predpol identifies and alerts police to areas in neighbourhoods where serious crimes are more likely to occur during a particular period. Researchers, activists and the media have all raised issues with this approach, citing that using historical data in this way creates a feedback loop, and argue that when police are focused to the idea that a particular neighbourhood is 'bad' instead of on the individual person before them, it changes the way they act towards that person based solely on their location. See: https://www.smithsonianmag.com/innovation/artificial-intelligence-is-now-used-predict-crime-is-it-biased-180968337/

With this in mind we have been developing lawmaps in a discrete set of legal domains. Fig. 5 describes the relationships on our pathway to developing legal AI. Lawmaps (which we will define formally in Section 3) are developed from legislation, precedent and procedural rules, all framed by expert input that illuminates flow, meaning, decision-making and reasoning processes. We believe lawmaps provide a logical structure which, when combined with relevant legal datasets and a common data model (CDM) for law, can support production of Legal AI.

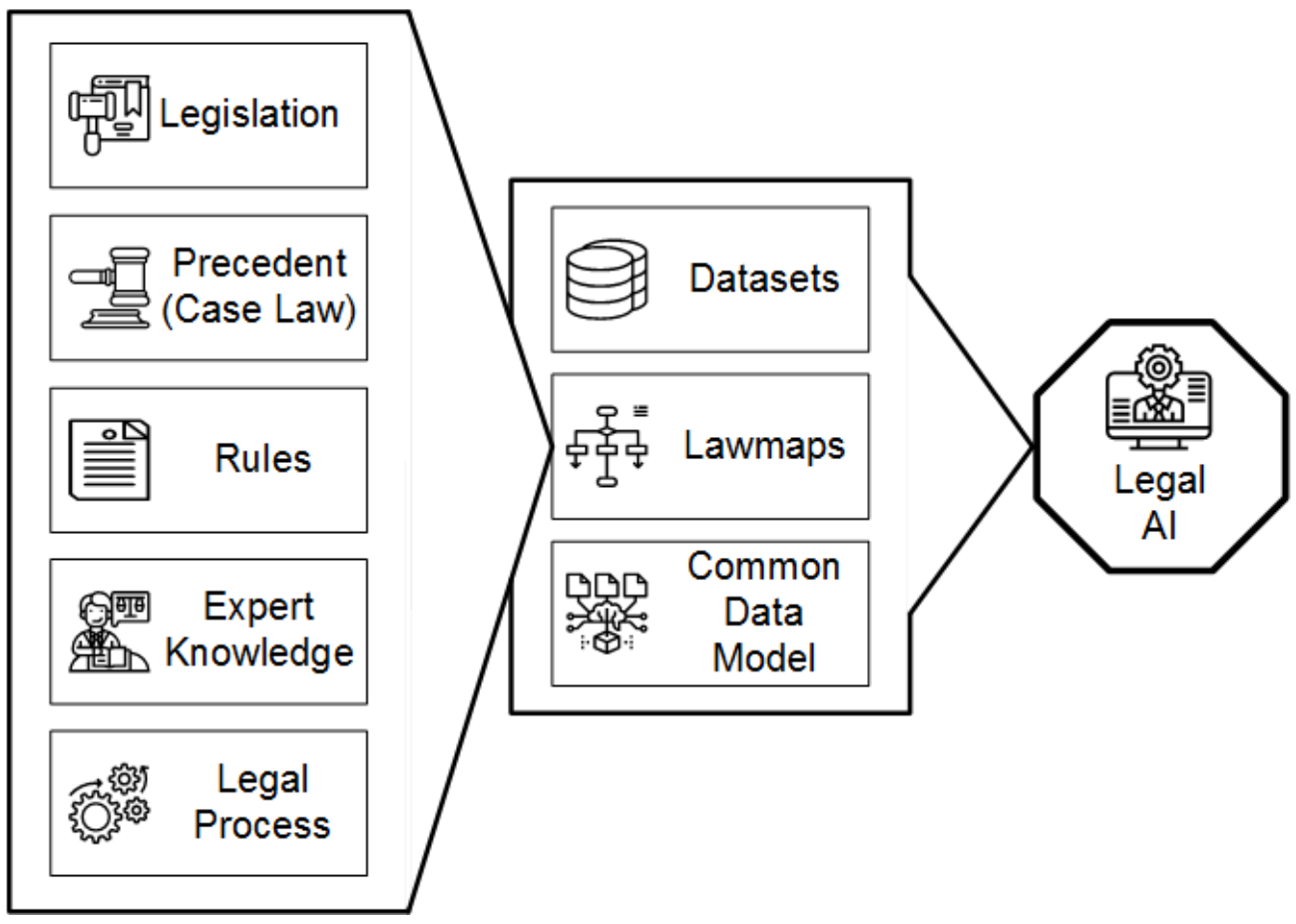

**Figure 5.** The pathway to Legal AI

## 3. Lawmaps: an approach for representing legislation and legal processes

Lawmaps grew from a need for infovis in the legal domain and out of the prior efforts of several of this works' authors to develop standard diagrams incorporating knowledge from clinical practice guidelines, published texts and clinical expertise in the medical domain. Known as *caremaps* (McLachlan et al, 2019), the resulting medical visualisations use a custom set of familiar unified modelling language (UML) elements to represent diagnostic and treatment activities, clinical decisions and the pathway a patient takes during the course of treatment for a given medical condition. Scaffolding a new infovis approach with an established presentation model that is familiar to the practitioner mitigated the need for extensive explanation and training of the representation framework, allowed the user to engage more freely with the knowledge content of the diagram, and resulted in an approach that practitioners agreed was simpler and easier to use than the text-based clinical guidelines and literature on which the caremap had, in part, been based (McLachlan et al, 2020b).

A recent review of process and flow diagram usage in legal literature published during the last two decades established that the *concept flow* (also known as flowchart or process map) was the most frequent visualisation in legal literature, and that UML was the single most recognisable representational language model (McLachlan & Webley, 2020). For this reason the lawmaps approach described in this work draws on the UML activity diagram, a common flow chart structure, for its elements, notation and visual appearance.

### 3.1 *Structure, elements and notation*

A lawmap contains a pathway, and each pathway presents as a sequence of elements, including: (a) functional entry and exit points at which the pathway is initiated or concludes; (b) activity nodes which represent functional components of legislation or the practitioners lawyerly effort; and (c) decision points which represent some legal decision to be made based on one or more criterion. The flow from one activity to another is illustrated with arrows. The pathway describes both: (i) the timing of necessary activities, that is, when they should occur and any dependencies required before performing the next activity; and (ii) the route or order of events identified as a result of the impact of evaluation of criterion at decision points. All elements and notation for lawmaps are described in Table 1.

### 3.2 *Lawyerly process lawmaps*

Each activity in the lawyerly process lawmap represents an activity that the lawyer undertakes in furtherance of a client's matter, including but not limited to: (a) taking instructions; (b) providing advice; (c) receiving or responding to correspondence; (d) investigation and consideration of legal matters; (e) generating briefs, written submissions or other legal documents; (f) recording resource expenditure and accounting to client; and (g) regulatory compliance activities. Additionally, an explanation associated with the activities, decision points and/or arrows may also be present. This could include an explanation for why the lawyer undertakes a particular activity, a description of the task to be performed, advice that should be given to the client, information that must be recorded, or correspondence to be generated. The example in Fig. 6 is extracted from a larger conveyancing lawmap[5]. It demonstrates a primary lawmap with nested lawmaps for each primary activity, along with examples for most explanation types.

[5] The complete conveyancing lawmap is available from:
http://www.mclachlandigital.com/lawvis/_include/img/work/full/OutlineConveyancingRegandUnregv3.png

**Table 1:** The lawmap content type, activities and decisions (Adapted from 28)

| | Element | Description | Notation |
|---|---|---|---|
| 1 | *Entry point* | Beginning of the lawmap | 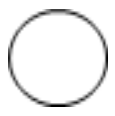 |
| 2 | *Exit point* | End of the lawmap | 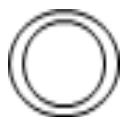 |
| 3 | *Activity* | A legal activity associated with or required by the legislation, case law, procedural rules or lawyerly process. | 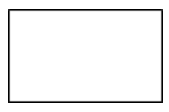 |
| 4 | *Nested Activity* | An activity that has its own underlying lawmap | 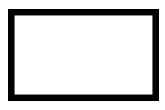 |
| 5 | *Decision* | A cognitive process of selecting a course of action that is associated with a legal question or consideration | 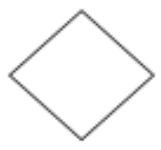 |
| 6 | *Nested Decision* | A decision that requires its own underlying lawmap | 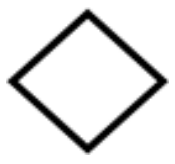 |
| 7 | *Flow* | Transition from one activity to another along the pathway | 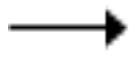 |
| 8 | *Multiple pathways* | Flow from an antecedent activity to a number of successors from which a decision point arises | 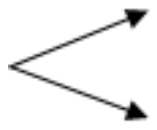 |
| 9 | *Decision Criterion* | Conditional values used to identify the path to be taken based on the legal decision being made |   |
| 10 | *Dependency or nested connection* | Connection between an activity in one pathway and either a dependency in another pathway, or a nested lawmap | 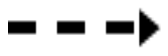 |
| 11 | *Multi-level lawmap connection* | Connection between a series of linked lawmaps | 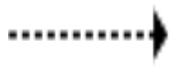 |

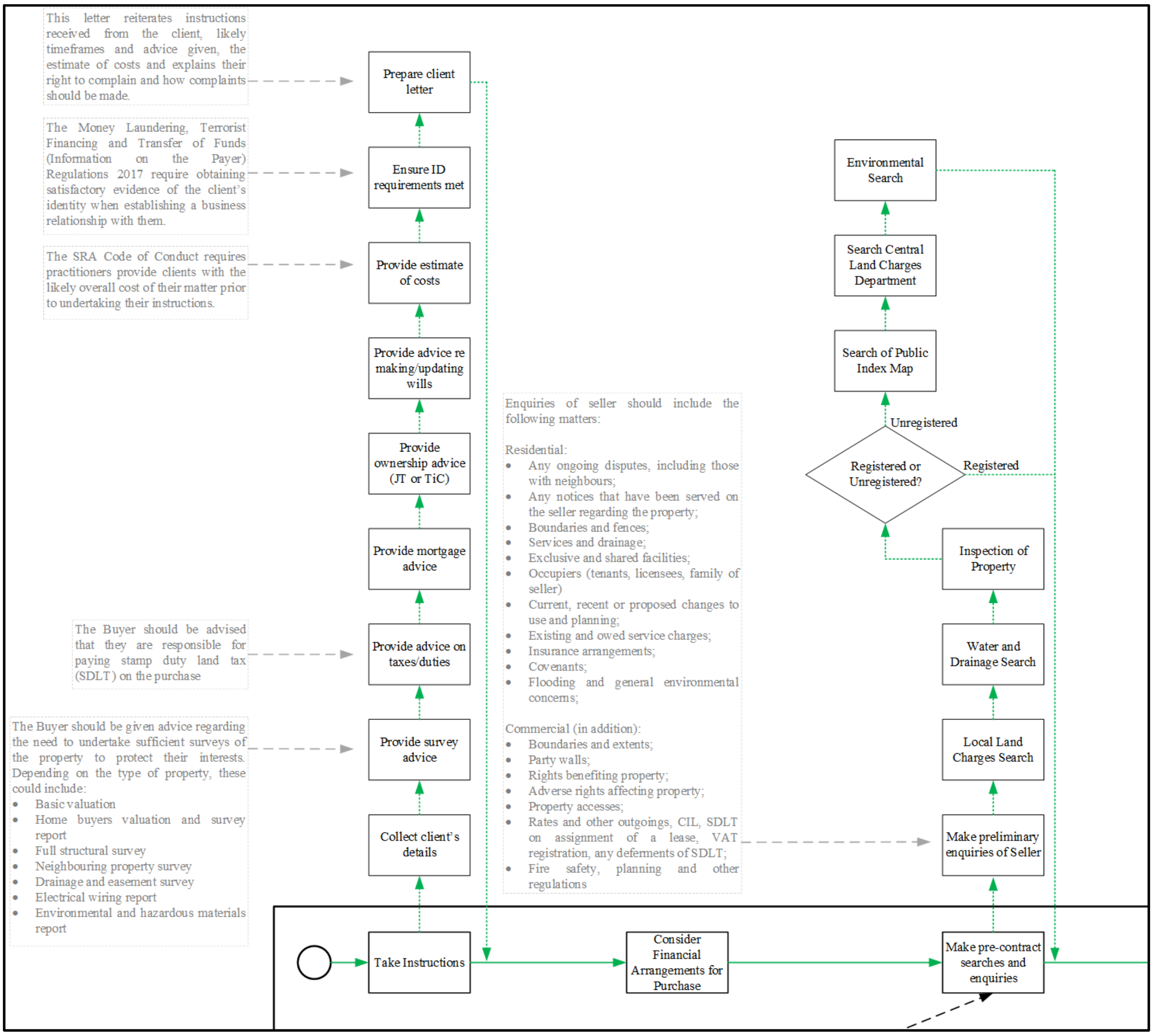

**Figure 6.** Extract of the multi-level Buyer pathway from conveyancing process lawmap

### 3.3 *Legislative lawmaps*

Each activity of the legislative lawmap represents either: (a) application of a specific section or subsection of the legislation; (b) a specific activity required by the legislation; or (c) the outcome from consideration of application of a section or subsection of the legislation (i.e. the outcome of decision points identified within the legislation). As with the lawyerly process caremaps there are a number of explanation breakouts that may be necessary. The example in Fig. 7 is extracted from the *Application for Interim Rent* process[6] found in Part II of the United Kingdom's *Landlords and Tenants Act 1954* and shows that these include references to the subsection of legislation from which the activity node, decision point or pathway has arisen, as well as references to relevant caselaw that establish or further refine that area of legislation.

[6] The complete Application for Interim Rent lawmap is available from: http://www.mclachlandigital.com/lawvis/_include/img/work/full/Leasesv0.4.png

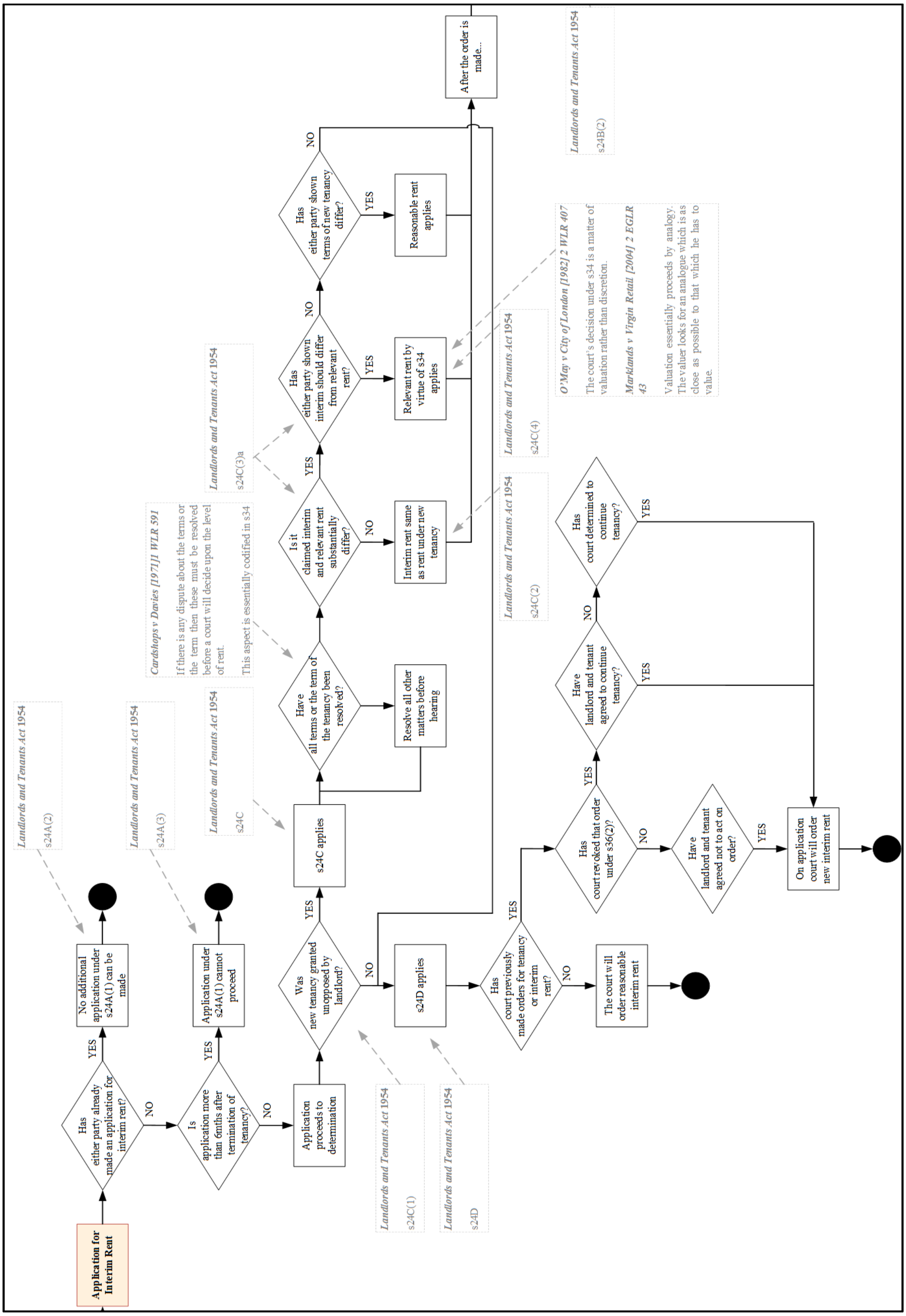

**Figure 7.** Extract of the Application for Interim Rent pathway drawn from the *Landlords and Tenants Act 1954*

## 4. Lawmap development

As shown in yellow at the apex of Fig. 8, the *first* phase in developing any lawmap is selection of the process to be mapped. The lawmap development lifecycle operates in the same way whether a legislative or lawyerly process is selected. The *second* phase, in green, locates source material and a practitioner with expertise relevant to the process under development. *Third*, in blue, the source materials are investigated to identify the overarching process flow which, using the source materials, standard constructs known as legal idioms and expert input, is analysed and broken down into component processes. The *fourth* phase in red draws on expert reasoning and evaluates the process flow and component processes to identify decision points at the apex of divergences described in the lawmap and the criterion relevant to each child path. Using an iterative *review and refine* approach, the *fifth* phase in purple draws on the outputs of all previous steps to design and draft the resulting lawmap.

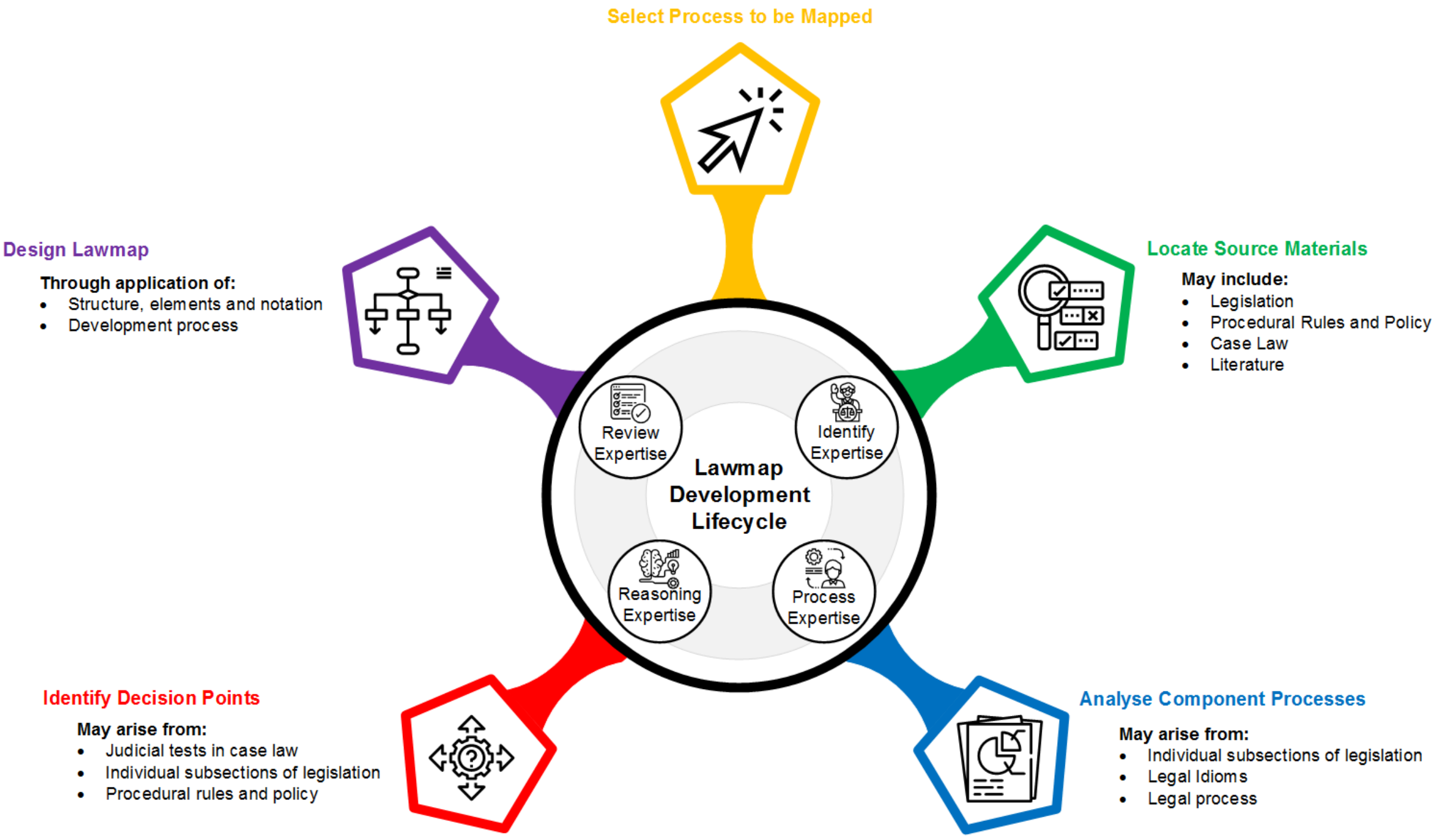

**Figure 8.** The Lawmap Development Lifecycle

As legislation is amended or new legal judgements reported, the lawmap may be updated through reprise of the lawmap development lifecycle with the updated source materials.

The design process described below assumes that the process being mapped has been identified and begins with collection of the source materials. The process for resolving the structure and flow of a lawmap follows an iterative per-node standard process described in points iii-vi below, with addition of the *Extract* step which is used to prepare legislation for direct application in that iterative standard process.

i. **Locate**
*Input:* The selected legislative or lawyerly process to me mapped
*Process:* Locate and collect relevant source materials to the process under investigation.

*Output:* A collection of relevant source material for use in development of the lawmap.
*Relates to:* The green phase in the Lawmap Development Lifecycle

ii. **Extract** (*for legislation*)
*Input:* Legislation from the source material collection
*Process:* Extract plain language logic from the law, making the legislation approachable and explaining its core elements, activities and application.
*Output:* A plain language process flow that describes the source legislation in simple terms.
*Relates to:* The blue phase in the Lawmap Development Lifecycle.

iii. **Identify**
*Input:* Output from the Locate and Extract steps.
*Process:* Identify the requirement for a node. Such requirement will be indicated from the plain language logic of legislation, conditions of a judgement or common law test, the structure of a legal idiom, or as a necessary step in procedural rules or lawyerly process being undertaken.
*Output:* A collection of identified nodes with analysis of their requirements, dependencies and operative rules.
*Relates to:* The blue phase in the Lawmap Development Lifecycle.

iv. **Distinguish**
*Input:* Output from the Identify step.
*Process:* Distinguish the type of node by its action and the effect it will have on the process flow. For example: where a node is an action or step to be undertaken and will be followed by another action or step, this will most likely be an activity node. Where there will be multiple pathways diverging from the identified node, this is most likely a decision point and necessitates identification of criterion to define the metric for selecting the individual divergent path to be taken.
*Output:* A complete set of nodes classified as activities or decision points.
*Relates to:* The red phase in the Lawmap Development Lifecycle.

v. **Sequence**
*Input:* Output from the Identify and Distinguish steps.
*Process:* The process to sequence a node requires identification of: (i) those activities that must be completed or that are necessary to the activity or decision to be undertaken in this node and which this node would have to follow; and (ii) any activities that depend on the outcome or completion of the activity or decision of this node and which this node must precede.
*Output:* The constructed lawmap.
*Relates to:* The purple phase in the Lawmap Development Lifecycle.

vi. **Traceability**
*Input:* Output of all previous steps.
*Process:* This step annotates the lawmap and enables traceability back to the legislation, case law or source material for nodes and the overall structure. Where the node has been resolved from a particular subsection of legislation, this should be indicated. In cases where a node arises from application, discussion or refinement of the law in judgement, the case law should be referenced with attention drawn to the legal reasoning that is required.
*Output:* Annotations for key nodes and pathways in the completed lawmap.
*Relates to:* The purple phase in the Lawmap Development Lifecycle.

## 5. Demonstration of the lawmap approach

It is beyond the scope of this paper to reproduce the complete process for developing both types of lawmaps: legislative or lawyerly. Instead, this section demonstrates a number of extract and component examples from each lawmap type using the terminology and activities described in the six-step design process above.

### *5.1 Lawyerly Process: Conveyancing*

*Locate:* The conveyancing process is normally outlined to law students as a list or table of activities as shown in Table 2[7].

*Identify* and *Distinguish:* Each component of the outline described in the source material was first analysed by the lawmap author and later reviewed with input from an expert conveyancing solicitor to identify the node, or nodes, and node type necessary to undertaking that activity as a step in the overall process.

*Sequence:* While a relationship between seller and buyer activities may be implied by their being colocated in the same row in the source table, relationships and dependencies between different activities were found to be more easily identified and located on the map when the basic process flow was represented in sequence for the expert. The first half of the conveyancing lawmap is shown in Fig. 9 with the seller and buyer pathways represented in corresponding colours to Table 2. Dependencies between Seller and Buyer activities identified by the expert are indicated with a black dashed arrow.

**Table 2:** General conveyancing: Outline of a simple conveyancing transaction (adapted from Abbey, *supra* note 33)

| **Seller** | **Buyer** |
|---|---|
| Take instructions | Take instructions and consider financial arrangement for purchase |
| Prepare and issue draft contract (sometimes with a draft purchase deed) | Make all necessary pre-contract searches and enquiries |
| Deduce (prepare and issue copies of the) title | Investigate title |
| Exchange contracts | Exchange contracts (arrange insurance) |
| Approve purchase deed | Make all necessary pre-completion searches |
| Prepare for completion | Prepare for completion |
| Completion | Completion |
| Post-completion procedures (including accounting to the client) | Post-completion procedures (including paying stamp duty land tax and title registration) |

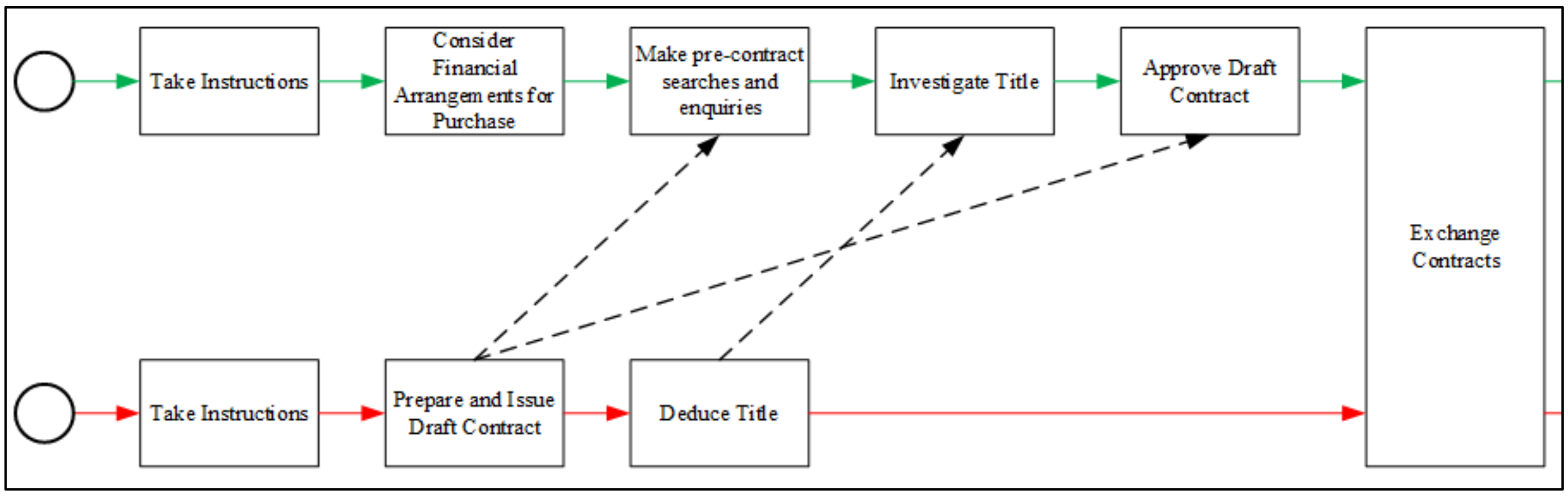

**Figure 9.** Extract of the overarching primary activities of the Conveyancing lawmap

[7] Adapted from: Abbey, R. & Richards, M. (2019). *Property Law*. Oxford University Press, UK, p 17.

As shown in Fig. 6 for the first few activity nodes of the Buyer pathway, each of the primary activity nodes in an overarching process flow will consist of either: (a) a nested lawmap: that is, a process flow that is undertaken to completion in order to complete the primary activity node; or (b) a component process represented in legislation, consistent with an established legal idiom or established from regulated procedure, practice rule or due diligence requirement. Often the individual steps of these component processes will be familiar to the expert practitioner, even though they may not be documented in textbooks or procedural practice notices with the same degree of granularity. It is these situations where expert elicitation is of primary importance.

*Traceability:* For the conveyancing lawmap we found a large number of activities that are undertaken by the experienced conveyancer arising from practice rules, regulatory policy and interaction between these and the requirements of legislation, for example: the *Money Laundering, Terrorist Financing and Transfer of Funds (Information on the Payer) Regulations 2017*. Traceability for the lawyerly process lawmap also includes annotations describing the content of specific work product that should be produced, such as the client engagement letter.

### *5.2 Legislative Process: The Landlords and Tenants Act 1954*

This example applies the lawmaps design process to an article of legislation; Section 24C of the *Landlords and Tenants Act 1954* (the Act).

*Locate:* Table 3 reproduces Section 24C of the Act.

**Table 3:** *Landlords and Tenants Act 1954*, s 24C.

| |
|---|
| **24C Amount of interim rent where new tenancy of whole premises granted and landlord not opposed** |
| (1) This section applies where— |
| (a) the landlord gave a notice under section 25 of this Act at a time when the tenant was in occupation of the whole of the property comprised in the relevant tenancy for purposes such as are mentioned in section 23(1) of this Act and stated in the notice that he was not opposed to the grant of a new tenancy; or |
| (b) the tenant made a request for a new tenancy under section 26 of this Act at a time when he was in occupation of the whole of that property for such purposes and the landlord did not give notice under subsection (6) of that section, |
| and the landlord grants a new tenancy of the whole of the property comprised in the relevant tenancy to the tenant (whether as a result of an order for the grant of a new tenancy or otherwise). |
| (2) Subject to the following provisions of this section, the rent payable under and at the commencement of the new tenancy shall also be the interim rent. |
| (3) Subsection (2) above does not apply where— |
| (a) the landlord or the tenant shows to the satisfaction of the court that the interim rent under that subsection differs substantially from the relevant rent; or |
| (b) the landlord or the tenant shows to the satisfaction of the court that the terms of the new tenancy differ from the terms of the relevant tenancy to such an extent that the interim rent under that subsection is substantially different from the rent which (in default of such agreement) the court would have determined under section 34 of this Act to be payable under a tenancy which commenced on the same day as the new tenancy and whose other terms were the same as the relevant tenancy. |
| (4) In this section “the relevant rent” means the rent which (in default of agreement between the landlord and the tenant) the court would have determined under section 34of this Act to be payable under the new tenancy if the new tenancy had commenced on the appropriate date (within the meaning of section 24B of this Act). |
| (5) The interim rent in a case where subsection (2) above does not apply by virtue only of subsection (3)(a) above is the relevant rent. |
| (6) The interim rent in a case where subsection (2) above does not apply by virtue only of subsection (3)(b) above, or by virtue of subsection (3)(a) and (b) above, is the rent which it is reasonable for the tenant to pay while the relevant tenancy continues by virtue of section 24 of this Act. |

*Extract:* Review of s24C using a plain language logic approach simplifies it into the listing presented in Table 4.

**Table 4:** Plain language listing of s24C of the Act

Where:

1. The Landlord has not opposed the granting of a new tenancy; or
2. The tenant has requested a new tenancy:
    a. By virtue of s26 of the Act;
        i. At a time when the tenant:
            1. Was in occupation
            2. Of the whole of the property; and
        ii. The landlord <u>has not</u> given effective notice:
            1. Under s26(6)
            2. Within two months of the tenant's request
            3. That he will oppose the grant of a new tenancy
            4. And the notice states grounds for opposition referenced in s30

Unless:

3. The landlord or tenant shows to the satisfaction of the court that:
    a. The interim rent should differ substantially from the relevant rent; or
    b. The new tenancy terms differ substantially from those of the existing tenancy:
        i. To the extent that the interim rent should differ substantially from the rent which the court would have determined.

In which case:

4. The court would determine the interim rent using s34 of the Act;

Otherwise:

5. The rent payable at the start of the new tenancy will be the interim rent.

*Identify* and *Distinguish:* This portion of the Act provides several relevant examples for describing the lawmap modelling approach. Analysis identified structurally necessary decision points that are shown in the lawmap extract presented in Fig. 10 and which arose out of consideration of points 1, 3, 3a and 3b of the plain language logic. While decision points are also evident from application of point 2 and its sub-points, in sequencing the lawmap we found these occur in the lawmap prior to the reproduced section. Review of other source material identified relevant case law in *Cardshops v Davies*[8] wherein the court found that where there is any remaining dispute regarding the terms or term of the tenancy then these matters must be resolved prior to commencing the interim rent hearing. The effect of this judgment is supported by s34 of the Act which in context assumes the existence of agreed terms by obliging the court to have *regard to the terms of the tenancy* when determining rent.

*Sequence:* It would be impossible for the court to determine an analogous valuation[9] for the tenancy where the terms or term of the tenancy remain irresolute. Sequencing this decision required identification of: (i) those activities whose completion would lead to consideration of terms; and (ii) any activities that depend on the presence of settled terms. First, consideration of terms would be unnecessary until, and only after, the tenant has sought a new tenancy which is unopposed by the landlord. Second, given that unresolved terms would prevent an interim rent application being brought before the court, the question of terms must have been resolved prior to those portions of s24C which, if in dispute, would necessitate adjudication of interim rent, and especially those that lead to a s34 rent determination.

[8] *Cardshop v Davies* [1971] 1 WLR 591.

[9] *O'May v City of London* [1982] 2 WLR 4007; The court's decision under s34 is a matter of valuation rather than discretion. *Marklands v Virgin Retail* [2004] 2 EGLR 43; Valuation essentially proceeds by analogy. The valuer looks for an analogue which is as close as possible to that which he has to value.

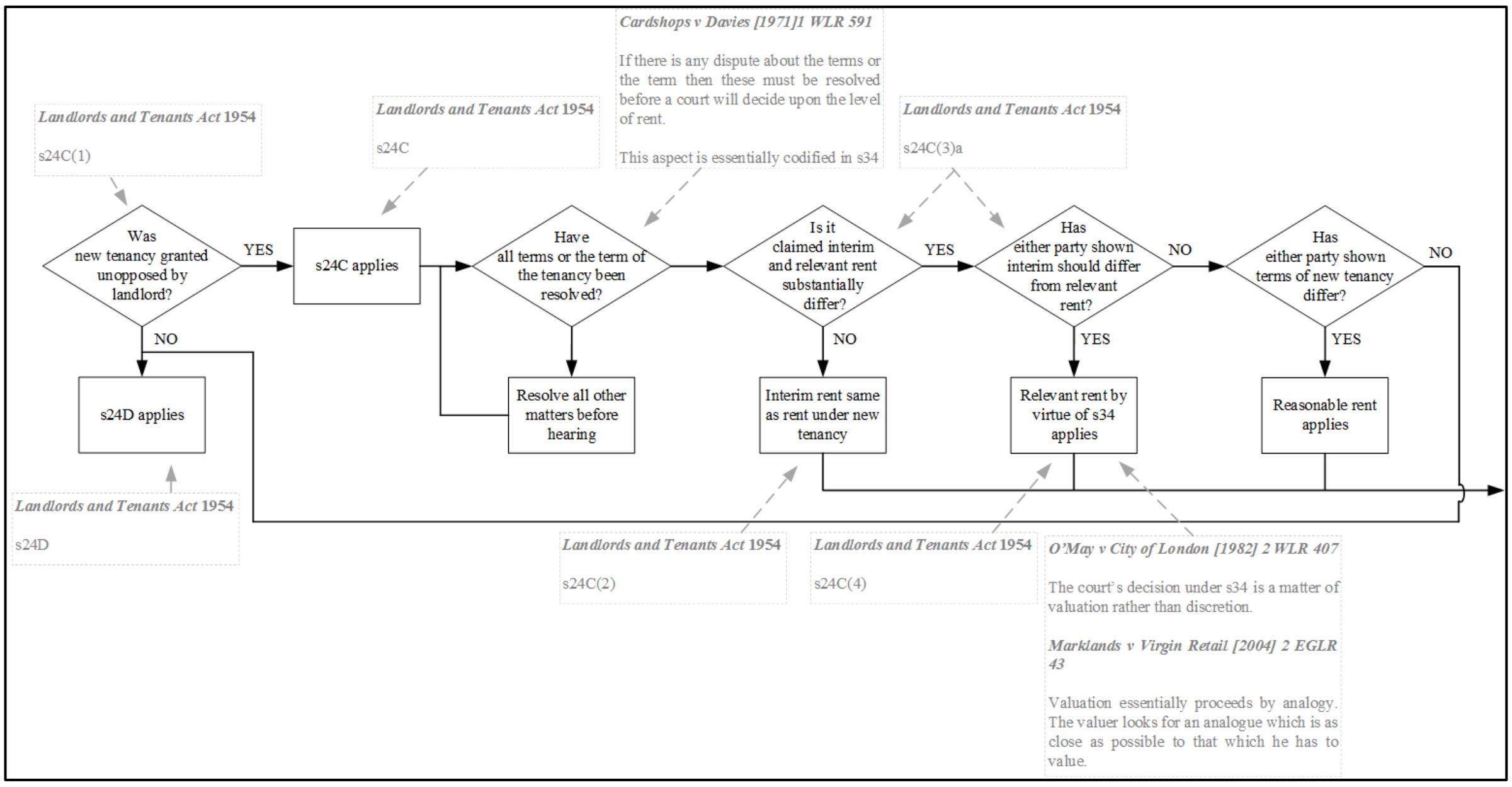

**Figure 10.** Extract of the *Application for Interim Rent* pathway drawn from s24C of the *Landlords and Tenants Act 1954*

*Traceability:* Annotations and explanations have been incorporated into the lawmap that call back to the source materials from which key nodes and pathways were resolved.

## 6. Discussion, Conclusions and Future Work

This paper presented *lawmaps*: an approach to visualising the implicit structure of legislation and the processes practitioners undertake when engaged in work on behalf of their clients. We are not the first group to promote the need, or even propose an approach, for infovis in the legal domain. However, we believe ours to be the first of a new breed of simple solutions capable of application across all facets, from legislation to practice; and capable of accelerating development of legal AI. We also believe the knowledge and skills to create lawmaps can be easily acquired by all, not just those in the legal domain.

It is not unusual in the legal domain for formulaic approaches like these to be dismissed. Some scholars caution against innovation or change in legal inculcation, practice or reasoning using approaches like the one proposed in this paper, which they describe as *cookie-cutter* methods (Tarr, 2004; Walton & Gordon, 2005). They do this often while simultaneously demanding broader and more enriched learning experiences for law students (Tarr, 2004). However, it becomes clear why many both within and outside the legal domain still consider the law to be unnecessarily complex and incomprehensible (Wagner & Walker, 2019)[10] when efforts that seek to bring clarity and ease of comprehension to the law such as lawmaps, or requirements to couch legislation in plain language, continue to be derided in this way (McLachlan & Webley, 2020)[11].

[10] The simple fact of the matter is that excessive, incomprehensible information is everywhere. Lawyers are particularly well acquainted with it. Unnecessary complexity and incomprehensibility complicate efforts to extract meaning from long, convoluted statutes.

One of the most iconic and relatable images of global legal systems is that of Themis[12], or as she is more commonly known, *Lady Justice* (Khorakiwala, 2020). Whether traditional as shown in Fig. 3, or contemporary as depicted in Fig. 4, she is recognised by her portrayal as a powerful woman with symmetry, grace and strong countenance and the items she carries in her hands: the *sword of retribution* grasped firmly in her right, and the balanced *scales of justice* suspended decisively below her left. However, where modern depictions diverge is on the iconographic representation of justice as blind. That is, whether or not Lady Justice is shown wearing a blindfold[13].

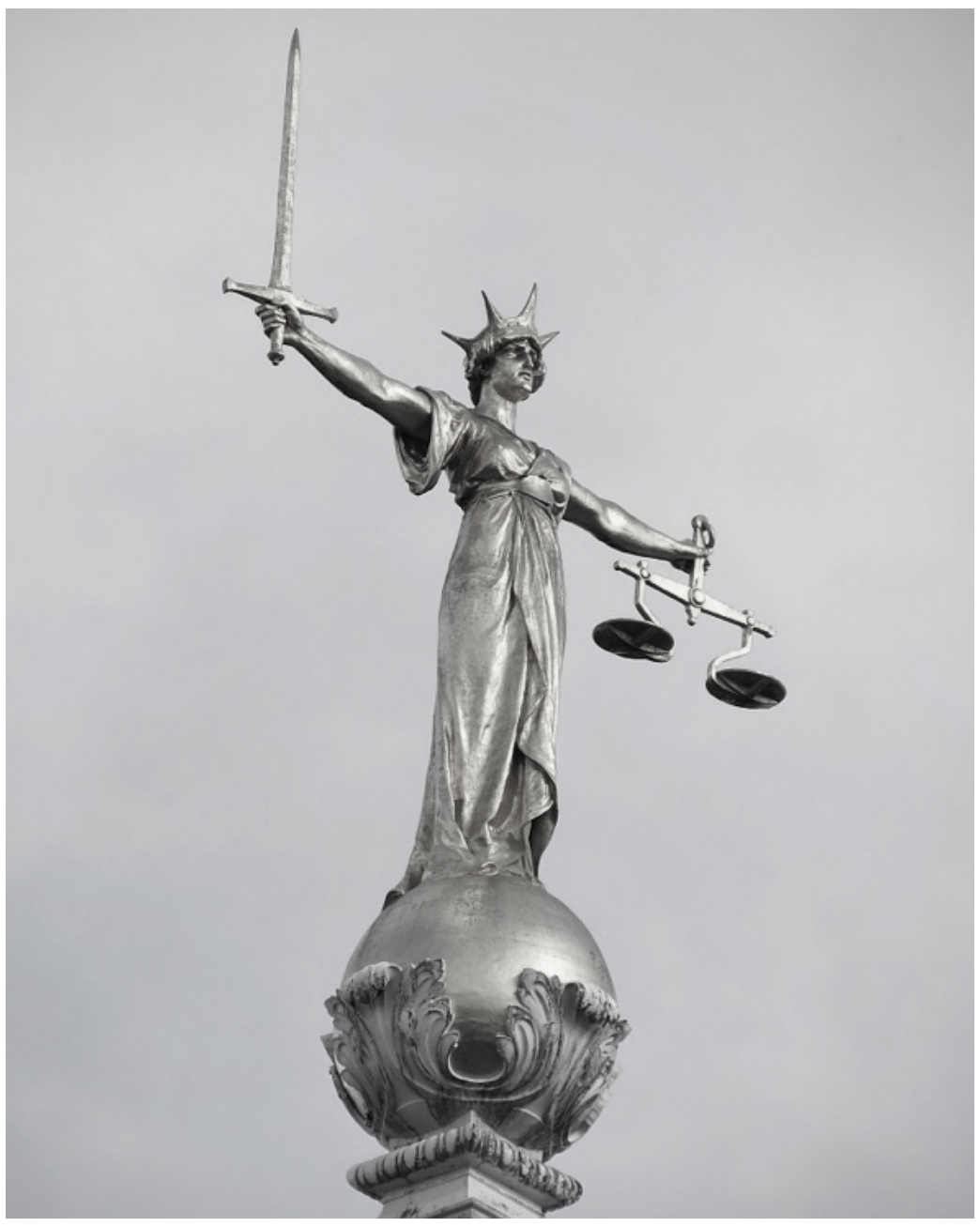

**Figure 3.** Lady Justice by Frederick Pomeroy (1906) on the Central Criminal Court of England and Wales, London, England (the *Old Bailey*)

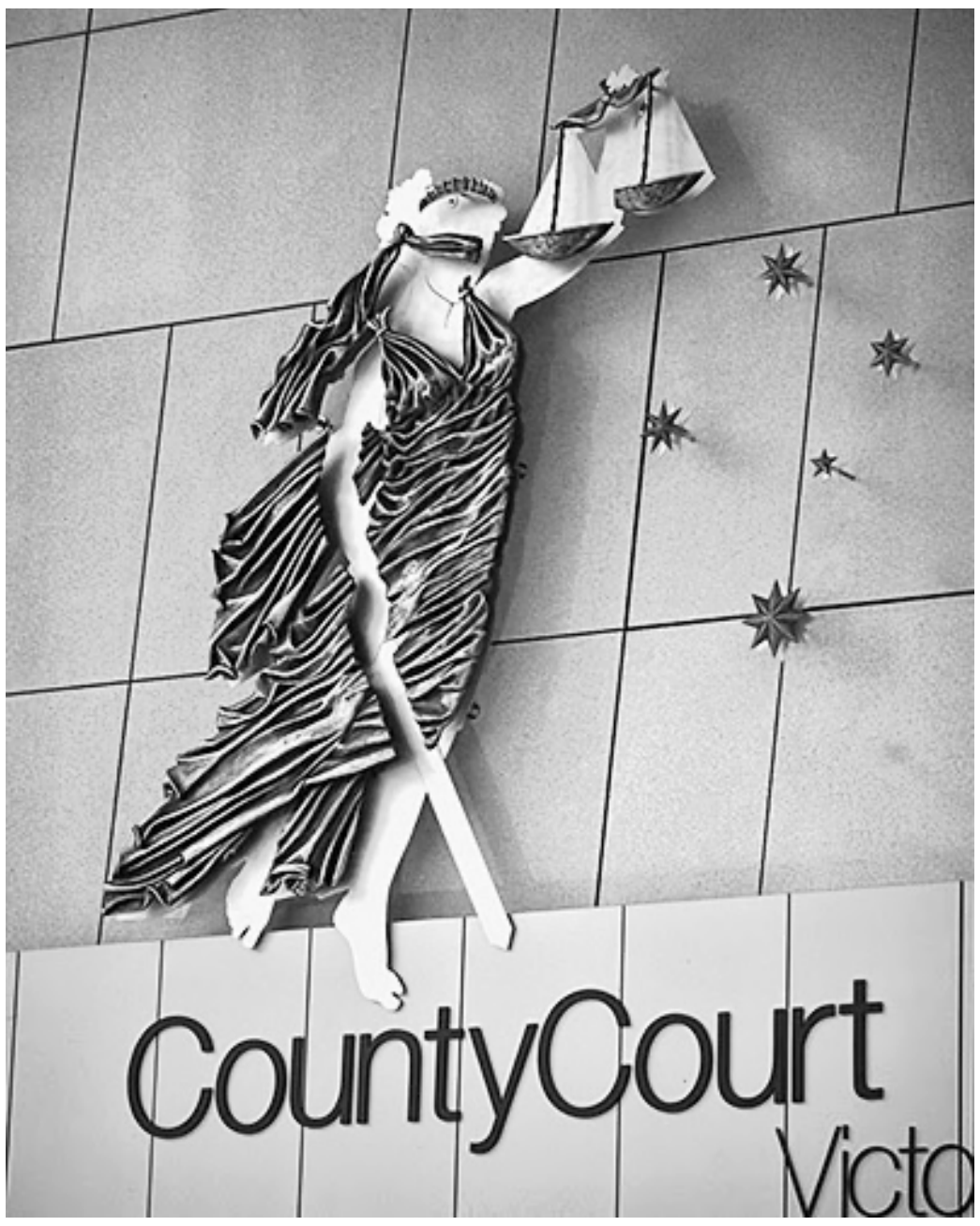

**Figure 4.** Lady Justice by William Eicholtz (2002) on the County Court of Victoria, Melbourne, Australia.

Lady Justice is said to have been blindfolded to guard against corruption of her objectivity and impartiality[14], and it is assumed that behind this veil she continues to see the law clearly in her mind's eye (Sisk, 2010). However, care should always be taken to ensure her blindness does not extend in both directions. While justice should be blind to those who come before the court, the right of those being judged to transparency of the judicial decision-making process means that courts have a duty to explain *why* and *how* a given ruling has come to be (Brennan, 1985). Explainability is also a key issue for Legal AI. Judges, lawyers, legal scholars and the general public have all decried the black-box nature of many Legal AI and espoused the need for those developing and using such systems to ensure they meet the legal requirement for Explainability

[11] Many of these approaches fall somewhere within the remit of what has become known as the *plain language movement*, which has at times been derided and misunderstood while being equally lauded for its capacity for sense and clarity.
[12] https://en.wikipedia.org/wiki/Themis
[13] As can be seen in the differences between Figures 3 and 4, depictions of Lady Justice wearing a blindfold are modern and not classical.
[14] The blindfold is said to represent her unfailing attitude that justice should be meted out without fear or favour. That the law should not regard appearance, wealth or power.

(Deeks, 2019; Olsen et al, 2019; Yu & Ali, 2019). One who is subject to a decision made by or with the assistance of AI has the same right to explainability as another receiving a jury or judge-alone verdict. Overall, the need for explainability is intended to avoid judgments made at random or in circumstances of uncertainty (Englich, Mussweiler & Strack, 2006). Lawmaps can support explainability for both human and algorithmic decision-makers by providing a visual chart of the structure of rules, legislation and precedent used in the decision-making process.

Prominent global issues during the first half of 2020 have resulted in significant job losses and an overwhelming economic downturn. Longstanding problems in the way legal services are delivered (Rhode, 2012), major reductions in funding and eligibility for legal aid (Flynn & Hodgson, 2017)[15], and an almost intentional lack of information regarding where people should go to get help when a legal issue arises all compound *access to justice* issues (Sudeall & Richardson, 2018). Often, complex situations arise either because a person lacks understanding about their rights and possible remedies in a given situation, or because they lack the knowledge for informed decision-making to understand whether they have taken the necessary actions to enact those rights (Sudeall & Richardson, 2018). Lawmaps are a simple tool for visualising legislation and legal processes that present otherwise complex topics in a way that is approachable to all. With inherent clarity, lawmaps are able to expose the implicit flow, knowledge and reasoning of a domain that has clung to incomprehensible text. Lawmaps represent one possible solution to mitigating *access to justice* issues by providing the untrained person with a visual primer to navigate their way through the intricacies of our legal system.

Recent articles provide diametrically opposed views on plain language law, AI and other technologies: while some welcome technology and promote new working models for lawyers, others engender fear through dire warnings that technology's sole purpose is wholesale replacement of the lawyer[16]. The latter may be the intention of some technology companies, however legal AI, democratising law and other approaches that produce *digital disruption* in the legal domain need not bring redundancy for the skilled practitioner. Legal training and practice must accept the inevitability of disruption and begin seeking approaches for transformation. If acknowledged and adopted within the legal domain, rather than replacing lawyers, approaches that may bring disruption can be used to: (i) streamline and improve delivery of legal services, (ii) reduce potential for errors and malpractice; and (iii) increase metaphorical foot traffic through the law firm door by providing those not trained in the law with a way to identify when, and why, they have an issue that should be brought to the expert practitioner.

Digitising lawmaps, merging them with data using a common data model and other structures like legal idioms will extend (Fig. 11) and may further simplify the pathway to developing Legal AI. They will be topics for future work.

---

[15] The most devastating cuts to legal aid in England and Wales began in 2013 following the introduction of the *Legal Aid, Sentencing and Punishment of Offenders Act 2012* which sought to reduce the legal aid budget by £350 million.

[16] Authors including Richard Susskind have promoted new working models for tech-adopting lawyers, while others like Gary Marchant have given dire warnings regarding their replacement by intelligent technologies - claiming that AI software has performed hundreds of thousands of hours of legal work and will displace as many as 100,000 lawyers from their jobs. See: Susskind, R., & Susskind, D. (2016). Technology will replace many doctors, lawyers, and other professionals. *Harvard Business Review*, *11*; *and* Nunez, C. (2017). Artificial intelligence and legal ethics: Whether AI Lawyers can make ethical decisions. *Tul. J. Tech. & Intell. Prop.*, *20*, 189; *and* Marchant, G. (2018) Artificial Intelligence and the future of Legal Practice. *Document Crunch*. https://www.documentcrunch.com/ai-news-artificial-intelligence-and-the-future-of-legal-practice.html

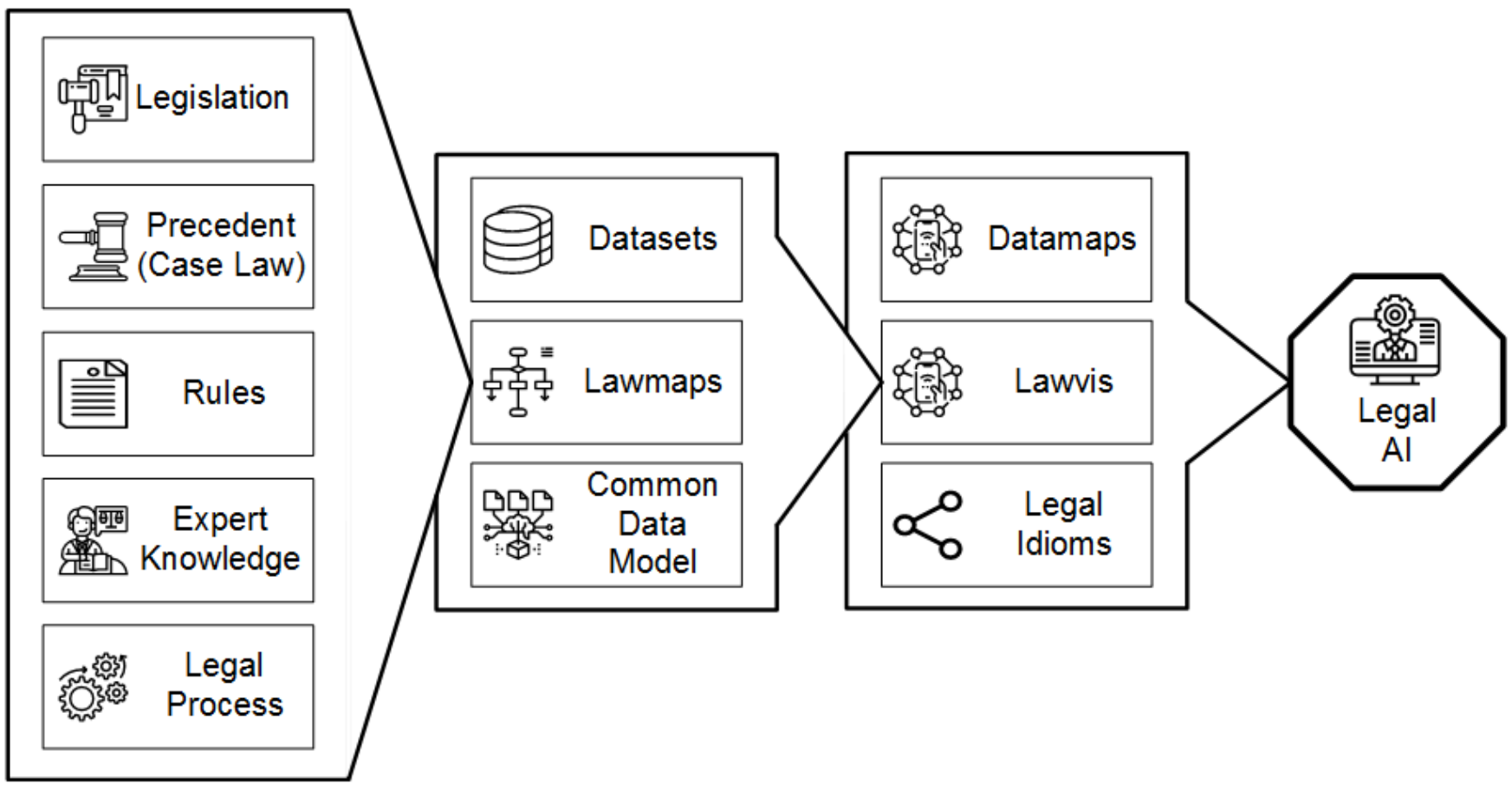

**Figure 11.** The extended pathway to Legal AI

We believe information visualisations like lawmaps will improve comprehension of the rights and duties imposed by law, and that they represent a significant step on the pathway to developing ethical, accurate and useable Legal AI. Lawmaps provide Lady Justice, and the rest of us, with easily comprehended visual knowledge of the interactions, inherent flow and processes of law that until now have remained obscured by her blindfold. As Gregory Sisk proposes (Sisk, 2010), *Lady Justice must lift her blindfold so that the visible contours of the law come into sharper relief.*